\documentclass[amssymb,prl,aps,twocolumn,amsmath,showpacs]{revtex4}

\usepackage[dvips]{graphicx}

\begin{document}

\title{Phase coherence of conduction electrons below the Kondo temperature}
\author{Gassem M. Alzoubi and Norman O. Birge}
\email{birge@pa.msu.edu}
\affiliation{Department of Physics and
Astronomy, Michigan State University, East Lansing, Michigan
48824-2320, USA}
\date{\today}

\begin{abstract}
We have measured the phase decoherence rate, $\tau_{\phi}^{-1}$ of
conduction electrons in disordered Ag wires implanted with 2 and
10 parts per million Fe impurities, by means of the weak
localization magnetoresistance.  The Kondo temperature of Fe in
Ag, $T_K \approx 4$~K, is in the ideal temperature range to study
the progressive screening of the Fe spins as the temperature $T$
falls below $T_K$.  The contribution to $\tau_{\phi}^{-1}$ from
the Fe impurities is clearly visible over the temperature range
40~mK -- 10~K.  Below $T_K$, $\tau_{\phi}^{-1}$ falls rapidly
until $T/T_K \approx 0.1$, in agreement with recent theoretical
calculations.  At lower $T$, $\tau_{\phi}^{-1}$ deviates from
theory with a flatter $T$-dependence.  We speculate that this
latter behavior is due to incomplete screening of the s=2 Fe
impurities by the conduction electrons.
\end{abstract}

 \maketitle

The Kondo problem is a paradigm many-body problem in condensed
matter physics.  In recent years, Kondo physics has been observed
in semiconducting quantum dots \cite{GoldhaberGordon} and carbon
nanotubes, while Kondo lattices play a major role in some
strongly-correlated materials \cite{HewsonBook}. The original
problem that motivated Kondo's 1964 paper \cite{Kondo1964} was the
increase in resistivity at low temperature of metals containing
magnetic impurities.  Although Kondo solved this puzzle, his
perturbation theory diverged in the limit of zero temperature.
Wilson \cite{Wilson1975} showed that, at temperatures far
below the Kondo temperature $T_K$, the magnetic impurity forms a
spin-singlet with the surrounding conduction electrons and behaves
as a non-magnetic scatterer with cross section given by the
unitarity limit.

The brief history given above might leave the impression that the
the behavior of dilute magnetic impurities in metals is completely
understood. That is, however, not the case. An important aspect of
the Kondo problem concerns the distinction between elastic and
inelastic scattering.  This distinction is extremely important in
the context of quantum transport and mesoscopic physics, where it
is known since 1979 that elastic scattering from static disorder
preserves quantum phase coherence, whereas inelastic scattering
destroys it.  The magnetic impurity contribution to the conduction
electron phase coherence rate, $\tau_{\phi}^{-1}$, was first
measured explicitly by two groups in 1987
\cite{PetersBergmann1987,vanHaesendonck1987}, and has received
renewed attention recently \cite{MohantyWebb2000, Schopfer2003} in
the context of the debate over zero-temperature decoherence in
disordered metals \cite{MohantyWebb1997, Pierre2003}.  Until very
recently \cite{Zarand2004,Micklitz2006}, however, there was no
theoretical expression for the temperature dependence of the
inelastic scattering rate valid for $T$ not too far below $T_K$,
and very little reliable data in that temperature range
\cite{Bauerle2005}.

The most reliable estimates of $\tau_{\phi}^{-1}$ come from
analysis of low-field magnetoresistance data in the context of
weak-localization theory.  In disordered metals without magnetic
impurities, $\tau_{\phi}^{-1}$ is dominated by electron-phonon
scattering at temperatures above about 1 K, and by
electron-electron scattering at lower $T$.  In the presence of
magnetic impurities, $\tau_{\phi}^{-1}$ contains the additional
contribution $\gamma_m$, which peaks at $T \approx T_K$.
In order to observe this peak in
$\gamma_m$ while keeping the magnetic impurity concentration low
enough to avoid interactions between impurities, one must choose a
system with $T_K$ below about 10 K; otherwise $\tau_{\phi}^{-1}$
is dominated by electron-phonon scattering.  In order to acquire
data far below $T_K$, however, it is important to keep $T_K$ as
high as possible. The optimal range for $T_K$ is a few Kelvins,
which is achieved with Fe impurities in Ag
\cite{WohllebenDaybell}.

We fabricated Ag wires of dimensions $L=780$~$\mu$m, $w =
0.1-0.2$~$\mu$m and $t = 47$~nm on oxidized Si substrates using
electron-beam lithography, thermal evaporation from a
$99.9999\%$-purity Ag source, followed by lift-off of the bilayer
resist.  All the wires studied were evaporated simultaneously to
ensure similar microstructure and resistivity. The 15 samples were
divided into four batches -- one that was kept pure, and three
that were implanted with 2, 6, and 10~ppm of Fe impurities,
respectively. SRIM simulations of the implantation at 70 keV
indicate that over 90\% of the implanted ions stay in the Ag
wires, with the rest going into the substrate
\cite{ImplantationUncertainty}.  As none of the 6~ppm samples
survived, we report data taken on a pure sample, and samples with
2 and 10~ppm Fe impurities.  Table I lists the sample parameters.
All samples were measured immersed in the mixing chamber of a
dilution refrigerator with filtered leads. Four-probe
resistance measurements were made using a lock-in amplifier with a
ratio transformer to improve sensitivity \cite{Pierre2003}. The
voltage drop across the sample was limited to $eV_s\lesssim k_BT$
to avoid heating the electrons.

\begin{table}[tbhp]
\begin{tabular}{c c c c c c c c c }
\hline\hline Sample &$L$ &$t$ &$w$ &$R$ &$D$ &$L_{so}$
&$L_{\phi}^{max}$&$c_{imp}$\\
&($\mu\mathrm{m}$)&($nm$)&($nm$)&($\Omega$)&($\mathrm{cm}^{2}/\mathrm{s}$)&($\mu
\mathrm{m}$)&($\mu\mathrm{m}$)&(ppm)

\\\hline\hline
Ag    & 780 & 47 & 130 & 3307 & 146 & 0.76& 14.6 &-\\
AgFe1 & 780 & 47 & 110 & 3890 & 146 & 0.86& 9.1  &2\\
AgFe2 & 780 & 47 & 185 & 2330 & 146 & 0.72& 4.1  &10\\\hline\hline
\end{tabular}
\caption{Geometrical and electrical characteristics of the
measured samples at 1.2~K. $L$, $t$, and $w$ are the sample
dimensions and $R$ is the resistance.  $D$ is the diffusion
constant determined from the resistivity and the Einstein relation
$\rho^{-1} =e^2D\nu_F$, with the density of states in Ag
$\nu_F=1.03\times10^{47} \mathrm{J^{-1}m^{-3}}$. $L_{so}$ is the
spin orbit length extracted from the fits of the magnetoresistance
to weak localization theory. $L_{\phi}^{max}$ is the maximum
coherence length measured at 40~mK. $c_{imp}$ is the implanted Fe
concentration. } \label{table1}
\end{table}

Raw magnetoresistance (MR) data for the three samples, at
$T=1.8$~K, are shown in the inset to Fig. 1. MR data for each
sample were fit using the following procedure. Because Ag has
moderate spin-orbit scattering ($\tau_{so} \approx 40$~ps for all
of our samples), the MR is positive at low temperatures, when
$\tau_{\phi} \gg \tau_{so}$.  At higher temperatures, when
$\tau_{\phi} < \tau_{so}$, the MR starts out positive but then
turns around at a field scale $B \approx 20$~mT.  Data at several
temperatures in this higher temperature range (which was different
for each sample) were first fit with three free parameters:
$L_{\phi} = (D \tau_\phi)^{1/2}$, where $D$ is the diffusion
constant, $L_{so} = (D \tau_{so})^{1/2}$, and the sample width,
$w$.  For each sample, these fits gave consistent values of
$L_{so}$ and $w$ over a broad temperature range
\cite{WidthUncertainty}.  We then fixed those values of $L_{so}$
and $w$, and fit the MR curves for all temperatures with
$L_{\phi}$ as the only free parameter.  Using the value of $D$
obtained from the resistance and sample dimensions, we finally
obtain $\tau_{\phi}$ as a function of temperature for each sample.

\begin{figure}[ptbh]
\begin{center}
\includegraphics[width=3.2in]{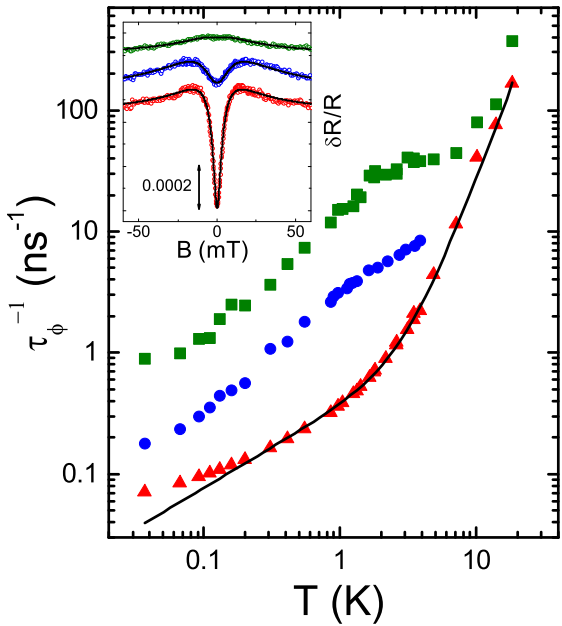}
\end{center}
\caption{(color online) Total decoherence rate vs. temperature for
a pure Ag sample ($\blacktriangle$), and for samples with
implanted Fe concentrations of 2~ppm ($\bullet$) and 10~ppm
($\blacksquare$). Inset: Raw magnetoresistance data at T=1.8 K for
the three samples, and fits to weak localization theory.}
\label{Fig1}
\end{figure}

Fig. 1 shows the decoherence rate, $\tau_{\phi}^{-1}$, obtained
from the MR data and fitting procedure described above, for the
three samples.  The $\tau_{\phi}^{-1}$ data for the pure sample
follow theoretical expectations above 200~mK, with a modest amount
of saturation at lower temperature \cite{Pierre2003}. Above
200~mK, the data are fit well (solid line) by a combination of
electron-phonon scattering ($\tau_{\phi}^{-1} \propto T^3$) and
electron-electron scattering ($\tau_{\phi}^{-1} \propto T^{2/3}$).
The prefactor of the $T^{2/3}$ term is equal to
0.36~ns$^{-1}K^{-2/3}$, in good agreement with the theoretical
value \cite{AAK}.  The additional contribution to
$\tau_{\phi}^{-1}$ from the Fe impurities, $\gamma_m$, is clearly
observable in the 2~ppm sample up to 4~K and in the 10~ppm sample
up to 10~K.  Above 10~K extraction of $\gamma_m$ from
$\tau_{\phi}^{-1}$ is not reliable, due to an apparent difference
in the magnitude of the electron-phonon scattering rate in the
10~ppm and pure samples.

The suppression of the weak localization magnetoresistance by
magnetic impurities was first discussed in
\cite{HikamiLarkinNagaoka1980} for the case of static impurity
spins.  Later, Fal'ko \cite{Falko1991} pointed out that the
impurity spins have their own internal dynamics given by the
Korringa rate, $\gamma_K \propto k_BT$.  Since spin-flip
scattering of conduction electrons and reorientation of the
impurity spins arise from the same processes, the ratio of their
rates is given by $\gamma_m/\gamma_K = n_s/{\nu_Fk_BT}$, where
$n_s$ is the density of magnetic impurities and $\nu_F$ is the
total density of states of conduction electrons at the Fermi
level. Below $T_K$ the Korringa rate saturates, so the ratio
becomes $\gamma_m/\gamma_K = n_s/{\nu_Fk_BT_K}$.  If $\gamma_K >
\gamma_m$, then the impurity spin configuration is randomized
during the time between spin-flip scattering events of the
conduction electrons. The contribution of magnetic scattering to
the total decoherence rate in the spin-singlet channel of the weak
localization magnetoresistance is given by:
\begin{equation}
\tau_{\phi}^{-1} = \tau_{in}^{-1} + 2\gamma_{m} \ \ \ \ \
\mathrm{for} \ \ \ \
 \gamma_m > \gamma_K. \label{staticspins}
\end{equation}
\begin{equation}
\tau_{\phi}^{-1} = \tau_{in}^{-1} + \gamma_m \ \ \ \ \;
\mathrm{for}\ \ \ \; \gamma_m < \gamma_K. \label{dynamicspins}
\end{equation}
where $\tau_{in}^{-1}$ is the dephasing rate due to
electron-electron and electron-phonon scattering. For our Ag
sample with 2~ppm Fe, assuming $T_K$ in the range 2-4~K (see
below), we find $\gamma_m/\gamma_K = 0.02 - 0.04 \ll 1$, hence we
should use Eq.~(\ref{dynamicspins}) to extract $\gamma_m$ from
$\tau_{\phi}^{-1}$.  Fig. 2 shows $\gamma_m$ for the 2~ppm and
10~ppm samples obtained from the $\tau_{\phi}^{-1}$ data in Fig. 1
using Eq.~(\ref{dynamicspins}), with $\tau_{in}^{-1}$ obtained
from the data on the nominally pure sample.  The data sets from
the 2~ppm and 10~ppm samples are consistent with each other when
we use Eq. (\ref{dynamicspins}) to analyze both of them.

\begin{figure}[ptbh]
\begin{center}
\includegraphics[width=3.2in]{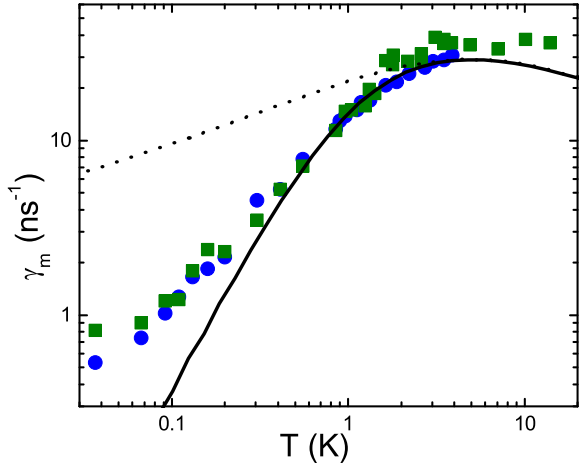}
\end{center}
\caption{(color online) Inelastic scattering rate due to magnetic
impurities for the 2~ppm ($\bullet$) and 10~ppm ($\blacksquare$)
samples.  Data for the 2~ppm sample are multiplied by 5.  The
solid line is the theoretical calculation of Micklitz \textit{et
al.} \cite{Micklitz2006} fit to the data for $T > 0.4$~K. The
dotted line is the Suhl-Nagaoka approximation for s=1/2.}
\label{Fig2}
\end{figure}

The criterion $\gamma_m/\gamma_K \ll 1$ is a necessary condition
for the theoretical approach of Micklitz \textit{et al.}
\cite{Micklitz2006} in their numerical renormalization group (NRG)
calculations of $\gamma_m$.  Those authors note, however, that Eq.
(\ref{dynamicspins}) is not strictly correct, because the
decoherence induced by electron-electron interactions \cite{AAK}
is not an exponential process.  Eq. (12) in \cite{Micklitz2006}
shows how to add the decoherence rate due to e-e interactions with
the rate due to all other processes including $\gamma_m$.  We have
compared the results of analyzing our data with Eq.
(\ref{dynamicspins}) and with Eq. (12) in \cite{Micklitz2006}, and
we find that for the 10~ppm sample the difference is negligible.
For the 2~ppm sample, using Eq. (12) in \cite{Micklitz2006} would
increase $\gamma_m$ by at most 15\% at the lowest temperature,
which is not significant on the logarithmic plot in Fig. 2.  The
solid line in Fig. 2 shows a fit of the NRG calculation from
\cite{Micklitz2006} to our data at temperatures above 0.4~K only.
Fitting to the 2~ppm (10~ppm) data alone gives $T_K = 4.8$~K
(5.4~K).  For comparison, the Suhl-Nagaoka approximation with $T_K
= 4.8$~K is shown as the dotted line. The data and fits shown in
Fig. 2 lead immediately to several striking conclusions. First, as
noted recently \cite{Bauerle2005}, the Suhl-Nagaoka approximation
does not come close to reproducing the temperature dependence of
$\gamma_m$ for $T < T_K$.  Second, the NRG theory of Micklitz
\textit{et al.} \cite{Micklitz2006} fits the data reasonably well
over the temperature range $T/T_K = 0.1 - 2$. And third, the NRG
theory deviates strongly from the data when $T/T_K < 0.1$.

According to \cite{Micklitz2006}, the maximum value of $\gamma_m$,
occurring at $T=T_K$, is $0.23 \times 4n_s/(\pi \hbar
\nu_F)$.  For 10~ppm of magnetic impurities in Ag, the theoretical
estimate is $\tau_{\phi}^{max} = 16$~$\mathrm{ns^{-1}}$, whereas
the data in Fig. 2 show a maximum value about twice as large. This
significant discrepancy may point to the inadequacy of the
spin-1/2 theory of \cite{Micklitz2006} to account for the large
spin (s=2) of Fe impurities in Ag.  On the other hand, previous
measurements of $\gamma_m$ in Ag samples implanted with Mn
impurities \cite{Pierre2003}, with s=5/2 and $T_K \approx 40$~mK,
were consistent with the theoretical estimate.  (Those
data were analyzed with the Suhl-Nagaoka approximation, which is
in close agreement with the theory of \cite{Micklitz2006} for $T >
T_K$.)  It thus appears that the inelastic scattering cross
section of Fe in Ag is roughly twice that of Mn in Ag.

\begin{figure}[ptbh]
\begin{center}
\includegraphics[width=3.2in]{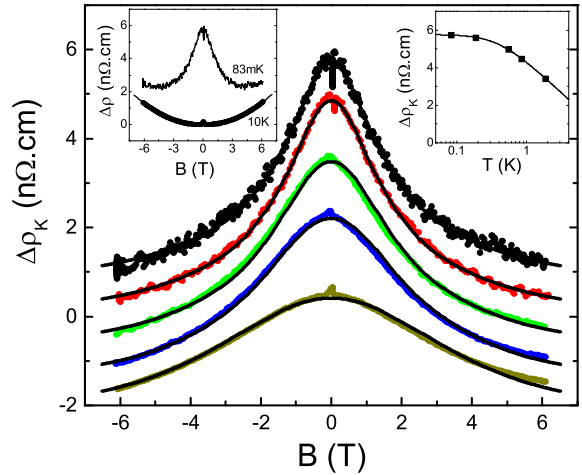}
\end{center}
\caption{(color online) Change in resistivity vs. magnetic field for
the 10~ppm sample at temperatures of 83, 188, 550, 860, and 1880~mK,
from top to bottom. The zero for the 83~mK data is chosen so that
the fit curve approaches $\Delta \rho \rightarrow 0$ for $|B|\gg
B_K$. Subsequent data sets are shifted downward by
0.75~$\mathrm{n}\Omega$cm for clarity. Solid lines represent the
global fit to the spin-1/2 NRG theory from \cite{Costi2000}, as
discussed in the text. Left inset: Raw $\Delta \rho(B)$ data at
83~mK and 10~K.  The latter shows the classical magnetoresistance
$\Delta \rho(B) \propto B^2$, which was subtracted from the raw data
sets to produce the curves in the main figure. Right inset: $\Delta
\rho(B=0)$ vs. temperature.} \label{Fig3}
\end{figure}

To further test the theory, we seek an independent estimate of
$T_K$ to compare with the value obtained from the fit to
$\gamma_m$ shown in Fig. 2.  The traditional method of determining
$T_K$ from the resistivity vs. temperature, $\rho (T)$, is not
reliable for our samples, because the electron-phonon scattering
contribution to $\rho (T)$ starts to grow significantly when $T >
8$~K, before the Kondo contribution has completely died out.
Instead, we analyze the high-field magnetoresistivity, $\rho(B)$,
shown at several temperatures in Fig. 3.  $\rho(B)$ data for
$T=53$~mK (not shown) are indistinguishable from those at $83$~mK,
indicating that the Kondo contribution to $\rho$ is close to the
unitarity limit at those temperatures.  We fit the $\rho(B)$ data
using NRG calculations from ref. \cite{Costi2000}.  For $T \ll
T_K$, these are nearly identical to the $T=0$ analytical solution
from \cite{Andrei1982}. The two-parameter fit to the 83~mK data
set gives the values $\Delta\rho_K\equiv\rho(B=0,T=0)-\rho(B\gg
B_K,T=0)=0.57$~$\mathrm{n}\Omega$cm/ppm and $B_K=1.2$~T, where $g
\mu_B B_K = k_B T_K$, with $\mu_B$ the Bohr magneton
\cite{T_K_definitions}. Using g=2, that in turn gives $T_K=1.6$~K.
Similar fits to the $\rho(B)$ data sets at higher temperatures,
keeping $\Delta\rho_K$ fixed, show that they can not be fit
consistently with such a low value of $T_K$.

To circumvent this problem, we have performed a global fit to all
the $\rho(B)$ data sets with only three parameters: $\Delta\rho_K$,
$B_K$, and $T_K$, where $B_K$ determines the magnetic field scale
over which $\rho(B)$ decreases, while $T_K$ determines the
temperature scale over which $\rho(B=0)$ decreases. The results of
this global fit are shown by the solid lines in Fig. 3, with the
values $B_K=1.36$~T, $T_K=2.96$~K, and
$\Delta\rho_K=0.58$~$\mathrm{n}\Omega$cm/ppm.  The value of $B_K$
corresponds to a Kondo temperature of 1.8~K.  The discrepancy
between the values of $T_K$ extracted from the field scale and from
the temperature scale highlights the fact that we are using a
spin-1/2 theory.  For free spins in a metal without Kondo effect,
the magnetoresistance is proportional to $-<s_z>^2$, where $<s_z>$
is the average spin polarization given by the appropriate Brillouin
function.  A comparison of the functions $-<s_z>^2$ for s=1/2 and
s=2 shows that the full width at half maximum of the s=1/2 function
is 1.85 times larger than for the s=2 function.  If a similar
relationship holds for the $\rho(B)$ curves in the Kondo regime,
then the value $B_K=1.2$T we found from the s=1/2 fit to the 83~mK
data would become $B_K\approx2.2$~T for s=2, which corresponds to a
Kondo temperature $T_K = 3.0$~K.  That is consistent with the value
of $T_K$ obtained from the temperature dependence of $\rho(B=0)$,
but still lower than the value $T_K \approx 5$~K obtained from the
fit to $\gamma_m$.

The experimental value $\Delta\rho_K=0.58$~$\mathrm{n}\Omega$cm/ppm
should be compared with the unitarity limit for s-wave scattering,
$\Delta\rho_K=4\pi \hbar n_s/n e^2
k_F=0.43$~$\mathrm{n}\Omega$cm/ppm for Ag, where $k_F$ is the Fermi
wavevector. (The value for d-wave scattering is five times higher.)
Our measured value of $\Delta\rho_K$ is larger than the theoretical
s-wave value, but less than the measured values in other Fe-noble
metal Kondo systems: 1.3~$\mathrm{n}\Omega$cm/ppm in Cu:Fe and
1.0~$\mathrm{n}\Omega$cm/ppm in Au:Fe \cite{Loram1970}.

The most serious discrepancy between theory and experiment is the
flattening of $\gamma_m(T)$ for $T/T_K<0.1$, shown in Fig. 2. The
behavior of the data is reminiscent of Eq. (3b) in
\cite{VavilovGlazman2003}, which describes inelastic scattering by
an underscreened Kondo impurity when $T \ll T_K$.  Since the
magnitude of $\gamma_m$ at $T=40$~mK is 40 times smaller than at
$4$~K, however, fitting the low-$T$ data with Eq. (3b) of
\cite{VavilovGlazman2003} would require a substantial reduction of
the prefactor.  The issue of screening for Fe in Ag is unclear. At
first glance, the five channels corresponding to the d-orbitals
should overscreen the s=2 Fe spin. In reality, however, the
different channels have different coupling constants, and the
impurity spin and orbital degeneracies are broken by crystal
fields and spin-orbit scattering.  Further theoretical work on
realistic models of spin-2 systems should help to resolve these
issues.

\textit{Note added}: We recently became aware of a related work
leading to similar conclusions \cite{Mallet2006}.

This work was supported by the NSF under grant DMR-0104178, and by
the Keck Microfabrication Facility under NSF DMR-9809688. We are
grateful to A.~Rosch and T.~Costi for sending us the results of
numerical calculations from \cite{Micklitz2006} and
\cite{Costi2000}, respectively.  We also thank C.~B\"{a}uerle,
L.~Glazman, S.D. Mahanti, R.~Ramazashvili, M. Vavilov, and G.~Zarand for
helpful discussions.


\begin{thebibliography}{9}
\bibitem{ImplantationUncertainty} Due to the small width of the wires, some ions may escape
from the sides of the wires.
\bibitem{WidthUncertainty} The values of $w$ obtained from the weak-localization
fits are 10-20\% smaller than those obtained from scanning
electron micrographs of the samples [10].
\bibitem{T_K_definitions} We use the definition of $T_K$ from [12], rather than from [20].
The latter is 2.05 times higher than the former.
\end{thebibliography}
\end{document}